\documentclass[pre,twocolumn,showpacs,amsmath,amssymb]{revtex4}

\usepackage{graphicx}
\usepackage{dcolumn}
\usepackage{bm}

\begin{document}

\preprint{}

\title{Power-law behavior in the power spectrum induced by 
Brownian motion of a domain wall}

\author{Shinji Takesue}
\email{takesue@phys.h.kyoto-u.ac.jp}
 \affiliation{Faculty of Integrated Human Studies\\ Kyoto University\\
 Kyoto 606-8501, Japan}
\altaffiliation[Address since April 1, 2003: ]{Department of Physics, Faculty
of Science, Kyoto University, Kyoto 606-8501, Japan}
\author{Hisao Hayakawa}%
\affiliation{%
Graduate School of Human and Environmental Studies\\
Kyoto University\\ Kyoto 606-8501, Japan
}%
\altaffiliation[Address since April 1, 2003: ]{Department of Physics, Faculty
of Science, Kyoto University, Kyoto 606-8501, Japan}
\author{Tetsuya Mitsudo}
\affiliation{Faculty of Integrated Human Studies\\ Kyoto University\\
 Kyoto 606-8501, Japan}
\altaffiliation[Address since April 1, 2003: ]{Department of Physics, Faculty
of Science, Kyoto University, Kyoto 606-8501, Japan}

\date{\today}

\begin{abstract}
We show that Brownian motion of a one-dimensional 
domain wall in a large but finite system 
yields a $\omega^{-3/2}$ power spectrum.
This is successfully applied
to the totally asymmetric simple exclusion process (TASEP) with open 
boundaries.  An excellent agreement between our theory and numerical results 
is obtained in a frequency range where the domain wall motion dominates and
discreteness of the system is not effective.

\end{abstract}

\pacs{05.40.-a, 02.50.Ey, 89.40.-a}
\maketitle

The asymmetric simple exclusion process (ASEP) is one of the
simplest models of collective transport of 
particles.\cite{spohn,D98,schutz}  
It can be interpreted
as a simplified model of traffic flow\cite{helbing,chowdhury} or 
as a model of granular flow.\cite{hayakawa} 
Since some exact solutions of ASEP can be obtained under given boundary
conditions, we may capture essential feature of more complicated systems 
such as traffic flows and granular flows.

In the study of the traffic flow or granular flow, 
the power spectrum $I(\omega)$ is used to investigate 
properties of local density fluctuations of vehicles or particles
and it commonly shows a power-law form $I(\omega)\sim\omega^{-\alpha}$.
For example,
in real observation of granular flow, it is confirmed 
that $\alpha=4/3$\cite{hayakawa,peng,moriyama}. 
On the other hand, the value of the exponent $\alpha$ has not yet fixed for 
traffic flows.\cite{musha,tadaki}
It is worthwhile mentioning that the structure factor (Fourie transform
of the power spectrum) in
the Rouse model, which is the simplest model of polymer dynamics, 
obeys $exp[-(t/\tau)^{1/2}]$ with the relaxation time $\tau$\cite{deGennes}.
Namely, the Brownian motion of structured material produces
$I(\omega)\sim \omega^{-3/2}$.

In this paper, we numerically calculate the 
power spectrum for the totally asymmetric simple exclusion process
(TASEP) which is the simplest ASEP and find that this model also exhibits
power-law $I(\omega)\sim \omega^{-3/2}$. 
This behavior can be explained by a
domain wall theory\cite{kolomeisky,santen}.

The TASEP is a continuous-time stochastic process on the one-dimensional
lattice which is defined as follows.  
Consider the set of lattice points $-L/2, -L/2+1,\dots,L/2$. 
Each site $i$ is either occupied by a particle ($\tau_i=1$) or empty 
($\tau_i=0$).  During an infinitesimal time interval $dt$, each particle 
hops to the right adjacent site with probability $dt$ if
the destination site is empty.  Moreover, in case of open boundaries,
a particle is injected to site $-L/2$ with probability $\alpha dt$ if the site 
is empty, and a particle is extracted from site $L/2$ with 
probability $\beta dt$ if the site is occupied.

The steady state of this model is exactly obtained for any 
parameter values $\alpha$ and $\beta$ and system size $L$ by the method of
matrix-product ansatz\cite{DEHP93}, which can also be applied to the more 
general ASEP\cite{Sasamoto}.
Moreover, in the steady state, the mean occupation
at site $i$, $\rho=\langle\tau_i\rangle$, and the current 
$J=\langle \tau_i(1-\tau_{i+1})\rangle$ can be calculated
in the large $L$ limit.
As the result, it turns out that 
there exist three phases depending upon values of
$\alpha$ and $\beta$ as follows. 
When  $\alpha<\frac{1}{2}$ and $\alpha<\beta$, the mean occupation is 
$\rho=\alpha$ and the current is $J=\alpha(1-\alpha)$.  This is called the low
density phase.  When  $\beta<\frac{1}{2}$ and $\beta<\alpha$, the mean 
occupation is $\rho=1-\beta$ and the current is $J=\beta(1-\beta)$.  This is
called the high density phase.  
When $\alpha>\frac{1}{2}$ and $\beta>\frac{1}{2}$, the mean occupation
is  $\rho=\frac{1}{2}$ and the current is $J=\frac{1}{4}$.  This is called
the maximum current phase.  The phase diagram is illustrated in Fig.\/ 1.

\begin{figure}
 \begin{center}
  \includegraphics[width=78mm]{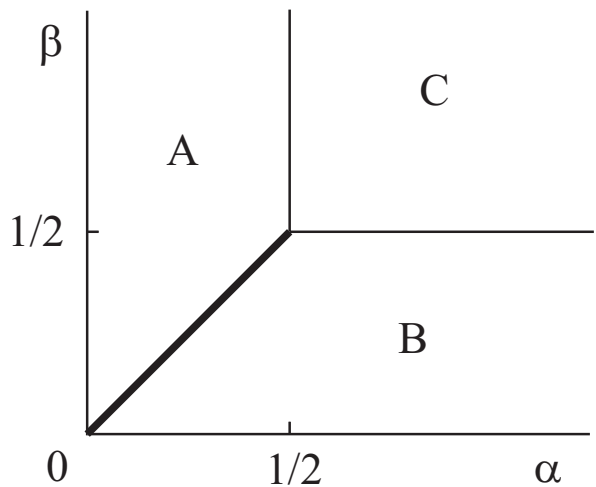}
 \end{center}
\caption{The phase diagram of the TASEP with open boundaries. 
Regions A, B and C mean the
 low-density phase, the high density phase and the maximum current phase,
 respectively.}
\end{figure}

The thick line in Fig.\/ 1, $\alpha=\beta<\frac{1}{2}$, is the
coexistence line, where we focus our attention in the following.  
In this case, there occurs a
domain wall between a left-hand region of the low density phase
and a right-hand region of the high density phase.  
It is known that the time-averaged density profile is a linear function 
of the position. This means that the domain wall can be anywhere with equal
probability as a result of its random walk.
This kind of kink motion can also be obtained from
a decouple approximation of ASEP where the equation of density profile
is reduced to the inviscid Burgers equation. It should be noted that the 
inviscid Burgers equation is a model equation of traffic
flow\cite{helbing,musha}.

The average motion and fluctuations of a domain wall is obtained 
as follows\cite{kolomeisky,santen}.
Assume that the system consists of a left
domain of particle density $\rho_{-}$ and flux $j_{-}$ and a right domain of
particle density $\rho_{+}$ and flux $j_{+}$. Then the drift velocity of the
domain wall is derived via the equation of continuity as
\begin{equation}
 V=\frac{j_{+}-j_{-}}{\rho_{+}-\rho_{-}}.
\end{equation}  
This is further interpreted as the result of random walk of the domain wall 
with hopping rates 
\begin{equation}
 D_{+}=\frac{j_+}{\rho_{+}-\rho_{-}},\quad
 D_{-}=\frac{j_-}{\rho_{+}-\rho_{-}},\quad
\end{equation}
for a move to the right ($D_{+}$) and the left ($D_{-}$).
In particular, on the coexistence line, the drift velocity vanishes and
the diffusion constant for the domain wall motion is given by
\begin{equation}
 D=\frac{\alpha(1-\alpha)}{1-2\alpha}.
\end{equation}

We carried out Monte Carlo simulations of the TASEP with open boundary
conditions of system size $L=200$.  
In our simulations, one Monte Carlo step corresponds to time interval $0.1$.
Parameter values are chosen to be $\alpha=\beta=0.2$, which is on the
coexistence line.
After discarding transient parts before a sufficiently large time $T_0$, 
sequences of site values $\{x_t=\tau_{y}(T_0+t)|t=0,1,\dots,T-1\}$ at a fixed 
position $y$ is recorded.  The length of the sequence is $T=2^{20}$.
Fourier components of the series is denoted by $\phi_n$.  That is,
\begin{equation}
 \phi_{n}=\frac{1}{T}\sum_{t=0}^{T-1}x_{t}e^{-i\omega_n t},
\end{equation}
where $\omega_n=\frac{2\pi n}{T}$ 
($n=0,1,2,\dots,$). 
Then the power spectrum $I(\omega)$ is computed by 
\begin{equation}
 I(\omega_n)=T\langle \left|\phi_{n}\right|^2\rangle,
\end{equation}
where the brackets mean averaging over 256 samples.

Figure 2 shows the simulation result for $y=0$, which lies in the center 
of the system. 
It is clearly observed that the power spectrum is propotional 
to $\omega^{-3/2}$ in the frequency range 
$5\times 10^{-5}\sim 2\times 10^{-3}$.  
In the following, we show that this power law behavior  
is induced by Brownian motion of the domain wall.

\begin{figure}[htbp]
 \includegraphics[width=86mm]{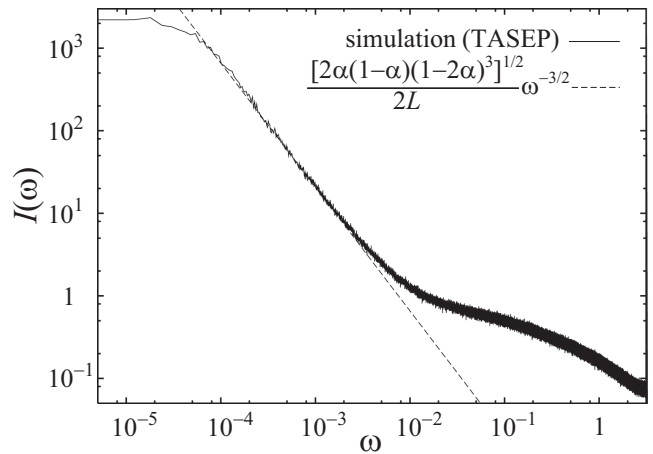}
\caption{Power spectrum fo time sequence $\{\tau_0(t)\}$ in the TASEP with
 open boundaries.}
\end{figure}

Now we consider a Brownian motion of a domain wall in an idealized form.
Let a domain wall be described by the step function $\theta(x)$ as
\begin{equation}
 \tau(x,t)=\theta(x-X(t)),
\end{equation}
where 
$X(t)$ is a Brownian motion in $[-L/2,L/2]$. 
The end points $-L/2$ and $L/2$ are assumed to be the reflecting boundaries. 
Then, the autocorrelation function of $\tau(y,t)$ is given by
\begin{equation}
 \langle \tau(y,t)\tau(y,0)\rangle =\int_{-L/2}^{y}dx_0\int_{-L/2}^{y}dx
 P_{\mathrm{st}}(x_0)P(x,t|x_0,0)
\label{corr}
\end{equation}
where $P_{\mathrm{st}}(x)$ is the stationary probability density for $X(t)=x$ 
and
$P(x,t|x_0,0)$ is the transition probability density.  In other words, if $D$
denotes the diffusion constant for $X(t)$, $P_{\mathrm{st}}(x)$ is the
stationary solution of the diffusion equation
\begin{equation}
 \frac{\partial P}{\partial t}=D\frac{\partial^2 P}{\partial x^2}
\end{equation}
with the boundary condition
\begin{equation}
 \frac{\partial P}{\partial x}\bigg|_{x=\frac{L}{2}}=
 \frac{\partial P}{\partial x}\bigg|_{x=-\frac{L}{2}}=0
\end{equation}
and $P(x,t|x_0,0)$ is the solution with initial condition 
$P(x,0|x_0,0)=\delta(x-x_0)$.  
The equation is easily solved and we find
\begin{equation}
 P_{\mathrm{st}}(x)=\frac{1}{L}
\label{P_st}
\end{equation}
and
\begin{eqnarray}
 P(x,t|x_0,0)&=&\frac{1}{L}
+\frac{2}{L}\sum_{n:\mathrm{even}}
 e^{-D\lambda_{n}^{2}t}\cos\lambda_{n}x_{0}\cos\lambda_{n}x\nonumber\\
& &+\frac{2}{L}\sum_{n:\mathrm{odd}}
 e^{-D\lambda_{n}^{2}t}\sin\lambda_{n}x_{0}\sin\lambda_{n}x,
\label{P_tr}
\end{eqnarray}
where $\lambda_n=\frac{\pi n}{L}$ with $n=1,2,3,\dots$.
Then, the integral in the right side of Eq.\/ (\ref{corr}) is carried out to
produce 
\begin{eqnarray}
\lefteqn{\langle\tau(y,t)\tau(y,0)\rangle=}\nonumber \\
& &\!\!\!\!\!\frac{(L+2y)^2}{4L^2}
+\sum_{n=1}^{\infty}
\frac{e^{-D\lambda_{n}^{2}t}}{L^2\lambda_n^2}
\left[1+(-1)^{n-1}\cos 2\lambda_n y\right].
\end{eqnarray}
This result can be generalized to negative $t$ by replacing $t$ by $|t|$ in
the above expressions.  

Because the power spectrum is 
equal to the Fourier transform of the autocorrelation function, we have
\begin{eqnarray}
 I(\omega)&=&\int_{-\infty}^{\infty}e^{-i\omega t}\langle\tau(y,t)\tau(y,0)
\rangle dt \nonumber\\
&=&\frac{\pi}{2}\delta(\omega)+\frac{2D}{L^2}
\sum_{n=1}^{\infty}
\frac{1+(-1)^{n-1}\cos 2\lambda_ny}{D^2\lambda_n^4+\omega^2}.
\end{eqnarray}
If the oscillatory part including factor $(-1)^{n-1}$ can be 
ignored and the sum over $n$ can be 
replaced by an integral, we can evaluate it as
\begin{eqnarray}
\frac{2D}{L^2}\sum_{n=1}^{\infty}
\frac{1+(-1)^{n-1}\cos 2\lambda_ny}{D^2\lambda_n^4+\omega^2}
&\simeq& \frac{2D}{\pi L}\int_{0}^{\infty}\frac{dx}{D^2x^4+\omega^2}
\nonumber\\
&=&\frac{\sqrt{2D}}{2L}\omega^{-3/2} 
\label{omega-3/2}.
\end{eqnarray}
Thus we have arrived at the $\omega^{-3/2}$ law.
It is remarkable that this result does not depend on position $y$.
Actually, as we will see later, $I(\omega)$ shows deviations from the above
result in small $\omega$ if $|y|$ is large.
This is because the oscillatory part cannot be ignored
in case of large $|y|$ and small $\omega$ even if system size $L$ is large.

Similar calculations can be done if the space is discrete.
Assume that a domain wall performs a 
continuous-time random walk on the one-dimensional lattice between two
reflecting walls located at $-\frac{L}{2}$ and $\frac{L}{2}$.  
Denoting hopping rate by $D$, we obtain
\begin{eqnarray}
\!\!\!\!\!\!\!\!\! I(\omega)&=&2\pi\left(\frac{[y]+1+\frac{L}{2}}{L+1}\right)^2\delta(\omega)
\nonumber \\
& &\!\!\!\!\!\!\!\!\!\!\!\!\!\!\!\!\!\!\!
+\frac{2D}{(L+1)^2}\sum_{m=1}^{L}
\frac{1+(-1)^{n-1}\cos\left(\frac{m\pi}{L+1}\left(2[y]+1\right)\right)}
{\omega^2+16D^2\sin^4\left(\frac{m\pi}{2(L+1)}\right)},
\label{theory1}
\end{eqnarray}
where $[y]$ denotes the largest lattice point that does not exceed $y$.
As in the previous case, if the oscillatory part can be ignored,
replacement of the sum in the right side by the integral yields for
$\omega\ne 0$
\begin{eqnarray}
I(\omega)&\simeq&
\frac{4D}{\pi(L+1)}\int_{0}^{\frac{\pi}{2}}\frac{dx}{\omega^2+16D^2\sin^2 x}
\nonumber \\
&=&
\frac{\sqrt{2D}\omega^{-3/2}}{2(L+1)}\left(
\frac{\frac{\omega}{4D}+\sqrt{1+\left(\frac{\omega}{4D}\right)^2}}{1+\left(\frac{\omega}{4D}\right)^2}
\right)^{\frac{1}{2}}.
\label{theory2}
\end{eqnarray}
In case $\omega\ll 4D$, this returns to the previous result.

\begin{figure}[htbp]
 \includegraphics[width=86mm]{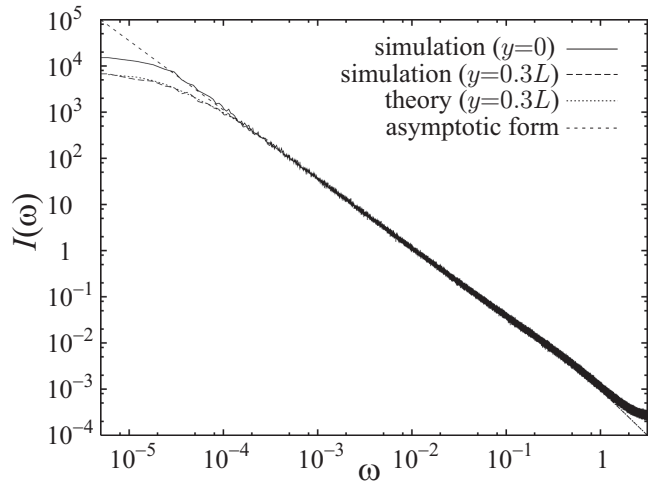}
\caption{Power spectra of the time sequences of site values
 $\{\tau(y,t)=\theta(y-W(t))|t=0,1,2,\dots\}$ for $y=0$ and $y=0.3L$,
where $W(t)$ means a random walk with hopping rate $D=0.1$.
The ``theory $(y=0.3L)$'' means the right-hand side of Eq.\/ (\ref{theory1})
 and the ``asymptotic form'' means the right-hand side of
Eq.\/ (\ref{theory2}).}
\end{figure}

To confirm the above results, we carried out numerical simulations.  
In the simulation, we deal with a system of length $L=200$ and 
let a domain wall carry out a random walk $W(t)$ with hopping rate $D=0.1$.  
Figure 3 shows numerically obtained power spectra in 
case $y=0$ and in case $y=0.3L$.
In case $y=0$,
the agreement between the theory and the numerical result is excellent.
Deviations from the power law are observed only in the very small 
frequency range $\omega \alt\pi^2 D/L^2\simeq 2.5\times 10^{-5}$, 
where discreteness of the spectrum $\{\lambda_n\}$ is eminent.
In case $y=0.3 L$, deviations extend to larger frequencies.
As mentioned earlier, this is because the oscillatory part cannot be ignored
if $|y|$ is large. 
In any case, however, the power law behavior is still clearly observed.

Now we return to the TASEP on the coexistence line.
Within the domain wall theory, $\tau_{y}(t)$ is approximated by
\begin{equation}
 \tau_{y}(t)=\alpha+(1-\alpha-\beta)\theta(y-X(t)),
\end{equation}
where $X(t)$ denotes the position of the domain wall at time $t$. 
Then, the prefactor of $\omega^{-3/2}$ becomes
$\sqrt{\frac{\alpha(1-\alpha)(1-2\alpha)^3}{2L^2}}$.  Actually,
the straight line in Fig.\/ 2 is depicted with this prefactor.  
The excellent agreement between the theory and the numarical result confirms
that the power-law behavior in the power spectrum is 
induced by the Brownian motion of a domain wall. 

\begin{figure}[htbp]
 \includegraphics[width=86mm]{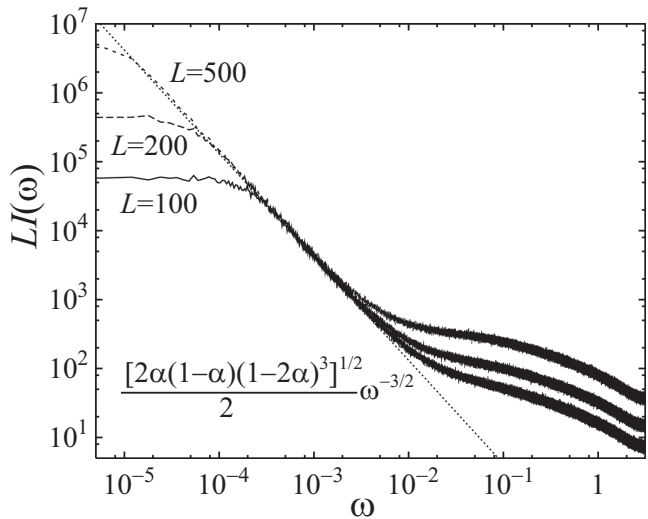}
\caption{Power spectra multiplied by system size for the TASEP of various
system sizes: $L=100$, $200$ and $500$.}
\end{figure}

The prefactor of the power law includes inverse of system size $L^{-1}$.  
This means that the power law in the power spectrum is a finite-size effect
and is observed only in large but finite systems.  
Figure 4 shows numerical results with various system sizes.  As system size
$L$ varies, the power-law part of the power spectrum $I(\omega)$ is scaled by
$L^{-1}$ as expected, while higher frequency part is not.  In fact, this part
is independent of system size $L$.  This means that the higher frequency part
represents local fluctuations.  

In summary, we have derived that Brownian motion of a domain wall
yields an $\omega^{-3/2}$ power spectrum.  This mechanism is successfully
applied to explain the power law in the power spectrum for local density
fluctuations in the TASEP.

It should be noted that the ASEP does not admit more than one domain wall.
In real traffic or granular systems, however, many domains can occur and 
interactions between them may be important.  It is a future problem to
investigate power spectra in such more complicated systems.

It is remarkable that the domain wall theory is quantitatively valid for TASEP.
This success may suggest the flutcuation of kink position is
characterized by a simple diffusion process. 
On the other hand, it is known that the current
fluctuation is not Gaussian\cite{johansson}.  We will have to clarify
this puzzled situation in the fluctuations of ASEP in the future.   

\begin{acknowledgments}
This work is partially supported by the Hosokawa Powder Technology
Foundation and the Inamori Foundation. TM and HH thank T. Sasamoto for
fruitful discussion.
\end{acknowledgments}

\end{document}